\documentclass[lettersize,journal]{IEEEtran}
\usepackage{amsmath,amsfonts}
\usepackage{mathrsfs}
\usepackage{algorithmic}
\usepackage{algorithm}
\usepackage{array}
\usepackage{titling}
\usepackage[caption=false,font=normalsize,labelfont=sf,textfont=sf]{subfig}
\usepackage{textcomp}
\usepackage{stfloats}
\usepackage{url}
\usepackage{verbatim}
\usepackage{float}
\usepackage{wrapfig}
\usepackage{graphicx}
\usepackage{cite}
\usepackage{svg}

\usepackage{subfig}
\usepackage{tabularx,booktabs}
\usepackage{multirow}

\usepackage{hyperref}
\usepackage{lipsum}
\usepackage[english]{babel}

\usepackage{soul}
\usepackage[normalem]{ulem}
\usepackage{amsmath}

\def\ss#1\end{{\color{blue}#1}\end}

\newcommand{\ra}[1]{#1}     

\newcommand{\rd}[1]{}      

\usepackage[switch,columnwise]{lineno}

\title{Machine-Learned Models for Power Magnetic Material Characteristics}

\date{}

\author{%
    Pawe\l{} Leszczy\'{n}ski\IEEEauthorrefmark{1}, 
    Kamil Kutorasi\'{n}ski\IEEEauthorrefmark{2}, 
    Marcin Szewczyk\IEEEauthorrefmark{1}, \IEEEmembership{Senior, IEEE},
    and Jaros\l{}aw Paw\l{}owski\IEEEauthorrefmark{3}
\thanks{Manuscript created Feb 16, 2024;
        \IEEEauthorrefmark{1}{Division of Power Apparatus, Protection, and Control, Faculty of Electrical Engineering, Electrical Power Engineering Institute, Warsaw University of Technology, Koszykowa 75 St., 00‑662 Warszawa, Poland}
        \IEEEauthorrefmark{2}{Department of Condensed Matter Physics, Faculty of Physics and Applied Computer Science, AGH University of Krakow, Reymonta 19 St., 30‑059 Kraków, Poland}
        \IEEEauthorrefmark{3}{Institute of Theoretical Physics, Wroc\l{}aw University of Science and Technology, Wyb. Wyspiańskiego 27 St., 50‑370 Wroc\l{}aw, Poland (\textit{Corresponding author: Jaros\l{}aw Paw\l{}owski}).}
}}

\begin{document}

\markboth{IEEE Transactions...,~Vol.~X, No.~X, \ldots}%
{}

\IEEEpubid{0000--0000/00\$00.00~\copyright~2024 IEEE}

\maketitle


\begin{abstract}
We present a general framework for modeling power magnetic materials characteristics using deep neural networks. Magnetic materials represented by multidimensional characteristics (that mimic measurements) are used to train the neural autoencoder model in an unsupervised manner. The encoder is trying to predict the material parameters of a theoretical model, which is then used in a decoder part. The decoder, using the predicted parameters, reconstructs the input characteristics. The neural model is trained to capture a synthetically generated set of characteristics that can cover a broad range of material behaviors, leading to a model that can generalize on the underlying physics rather than just optimize the model parameters for a single measurement. After setting up the model, we prove its usefulness in the complex problem of modeling magnetic materials in the frequency and current (out-of-linear range) domains simultaneously, for which we use measured characteristics obtained for frequency up to $10$ MHz and \rd{current}\ra{H-field} up to \ra{saturation}\rd{$27$ A}.
\end{abstract}


\section{Introduction}
\IEEEPARstart{M}{agnetic} components, such as inductors and transformers, are vital in power electronics devices requiring dedicated modeling efforts due to specific working conditions and underlying physical principles. No fully satisfactory first-principle models have yet been proposed for power magnetic materials, and current\ra{ly} efforts are underway to reproduce magnetic materials characteristics using machine learning frameworks~\cite{li2023magnet_1,serrano2023magnet,li2023magnet_2}. After developing a theory that describes the underlying physics, one comes to the stage where it is necessary to choose the parameters of the model, most often adjusting them to obtain a correspondence between the model characteristics and the measurements. Depending on the model complexity, a typical approach is to optimize parameters by curve fitting using least-squares or one of the vast variety of optimization algorithms, such as gradient-based methods\cite{grad}, simulated annealing\cite{sa}, particle swarm optimizations\cite{pso1,pso2}, or evolutionary algorithms\cite{evo,ga}. Regardless of the algorithm chosen, optimization is always the selection of a parameter for a given set of measurement data.

With the advent of big data, deep learning, i.e. paradigm that uses deep neural networks (NNs), revolutionary breakthroughs in many conventional machine learning and pattern recognition tasks are a fact, such as computer vision or signal processing~\cite{Chai2021,Alam2020}. The
accompanied data-driven materials modeling approach~\cite{selvaratnam2021,voyles2017,schutt2019,chen2019}, including data-driven magnetic materials modeling~\cite{li2023magnet_1,serrano2023magnet,li2023magnet_2}, also provides significant shift in materials science methodology from simple trial-and-error routine to intelligent discovery of new materials with demanding properties~\cite{Li_2022, Fuhr2022, Choudhary2022, Bonatti2021, damewood2023}.

In this paper, we propose a similar approach but for learning \textit{any} analytical model of a material in terms of intelligent fitting of its parameters. \ra{The only requirement for the model is that it must be differentiable to allow the fitting error to be passed back (backpropagated) during the NN training}. In the following, we train a machine learning model composed of a deep NN to be able to select parameters of a given theoretical model over a wide range of material characteristics at the same time, not just to single measurement data, which from machine learning perspective would be just overfilling of this single characteristic. \rd{Our goal is to build a model capable of generalizing over a variety of characteristics than a single one, which also means that the NN is able to capture the physics behind the theoretical model.} \ra{Our aim is to construct a model capable of generalizing across various features, not just one, which also implies that the neural network can capture the physics underlying the theoretical model.}

\ra{Neural networks have been applied for modeling core losses or hysteresis loops of power magnetics (e.g., nanocrystalline rings) using simple few-layer perceptrons~\mbox{\cite{corednn1, corednn2, corednn3, corednn4, corednn5, corednn6, corednn7}}, or recurrent NN~\mbox{\cite{corernn}}. Also, recently introduced \textit{MagNet} platform~\mbox{\cite{magnet1,magnet2}} proves that NN can be utilized to effectively learn B-H loop of magnetic material via more advanced autoencoder architectures, similar to ours.}

\ra{Nanocrystalline rings are currently used in many power and electronic devices~\mbox{\cite{ferch2013application}} to meet the requirements imposed on the devices with respect to electromagnetic compatibility by suppressing electromagnetic interferences and insulation coordination by suppressing overvoltages occurring in these devices, in particular, overvoltages occurring in Gas-Insulated Switchgear \mbox{\cite{riechert2012mitigation}}).}

\section{Model and data generation}
\rd{To show the capability of the proposed approach, we will focus on a simple but very useful phenomenological model of magnetic materials in a wide frequency range capturing linear but also nonlinear regimes\mbox{\cite{abb2016}}}.Various approaches to modeling magnetic materials in nonlinear conditions were proposed, including famous Jiles-Atherton (JA) model~\cite{JA1984,Liorzou2000}, models based on Finite Element Method (FEM) that can describe local spatial dependence of magnetization~\cite{Perkkio2018,Hoffmann2016}, or lumped element equivalent circuits (LEEC)~\cite{leecore}, suitable for direct implementation in SPICE or EMTP simulators~\cite{spice,dommel1974computation}.

\ra{To show the capability of the proposed approach, we will focus on a phenomenological model of magnetic material in a wide frequency range capturing both linear and nonlinear \ra{H-field} regimes}. Here, we will use LEEC representing impedance characteristics of magnetic rings in both frequency and current domains, introduced in~\cite{abb2016}, and further analyzed in terms of its parameters optimization~\cite{bib:NMMcircuitSimulation}, also using AI methods~\cite{PawlowskiJ2022Mnmo}.
\rd{
Nanocrystalline rings are currently used in many power and electronic devices~\mbox{\cite{ferch2013application}} to meet the requirements related to electromagnetic compatibility by suppressing electromagnetic interferences occurring in these devices and the requirements of insulation coordination by suppressing overvoltages occurring in these devices, in particular, overvoltages occurring in Gas-Insulated Switchgear.} \rd{The problem of overvoltage suppression in GIS is important in the case of high-voltage GIS starting from 550 kV and ultra-high voltage GIS up to 1100 kV~\mbox{\cite{riechert2012mitigation}}, in which fast-changing overvoltages (with frequencies of several MHz) overlap with slow-changing waveforms (with frequencies of 50/60 Hz), which are approximately constant. The significance of these conditions comes from the fact that the peak values of these overvoltages determine the design of the insulation systems of the GIS devices.}

\begin{figure*}[t]   
    \centering
            \begin{tabularx}{\textwidth}{ p{-2.6cm} l  p{-2.6cm}  l }
    (a) &   & (b) &  \\  
         & \includegraphics[width=0.455\textwidth, trim={1.1cm -0cm 0 0.1cm}]{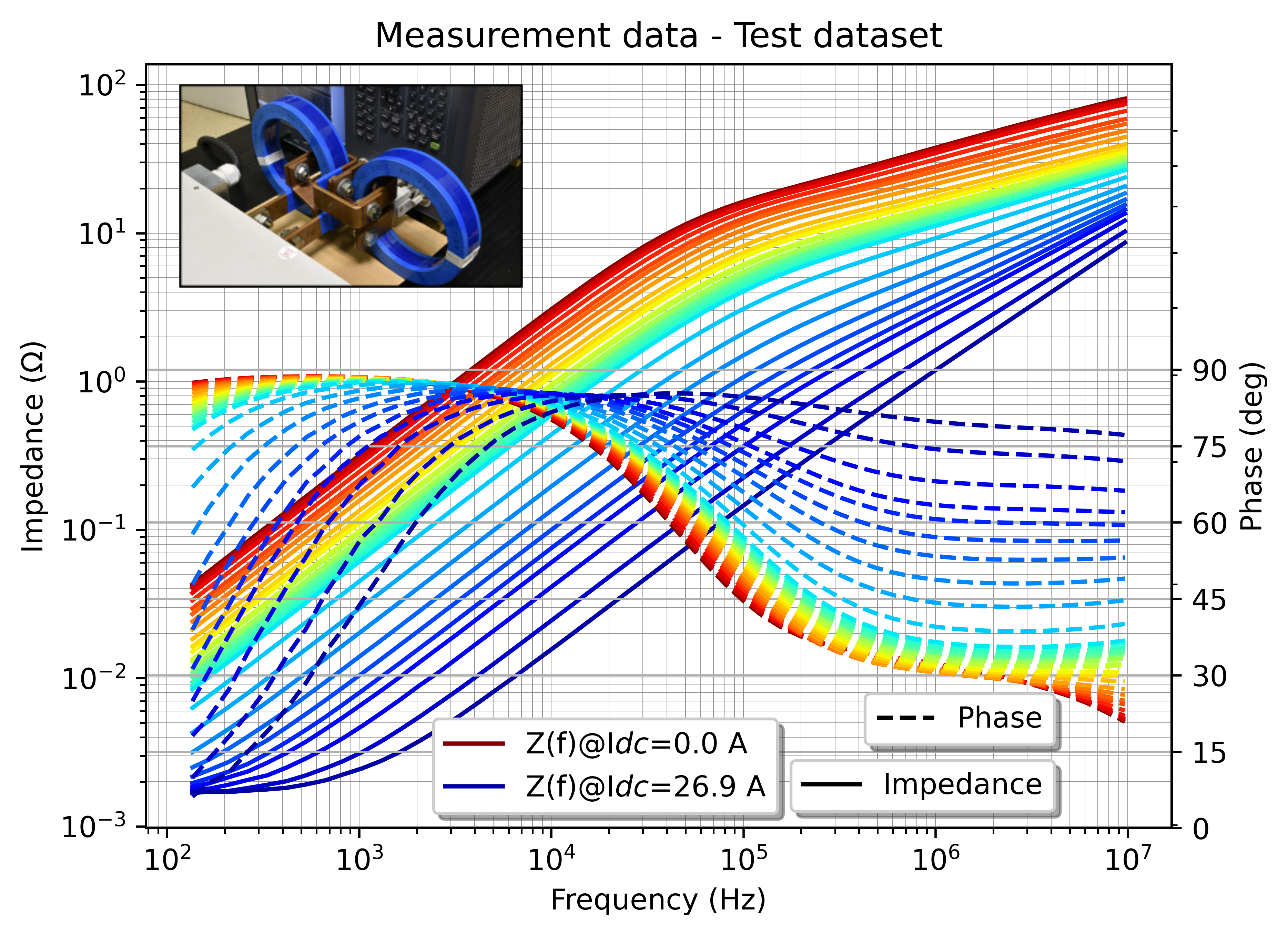} & &\includegraphics[width=0.455\textwidth, trim={1.1cm -0cm 0 0.1cm}]{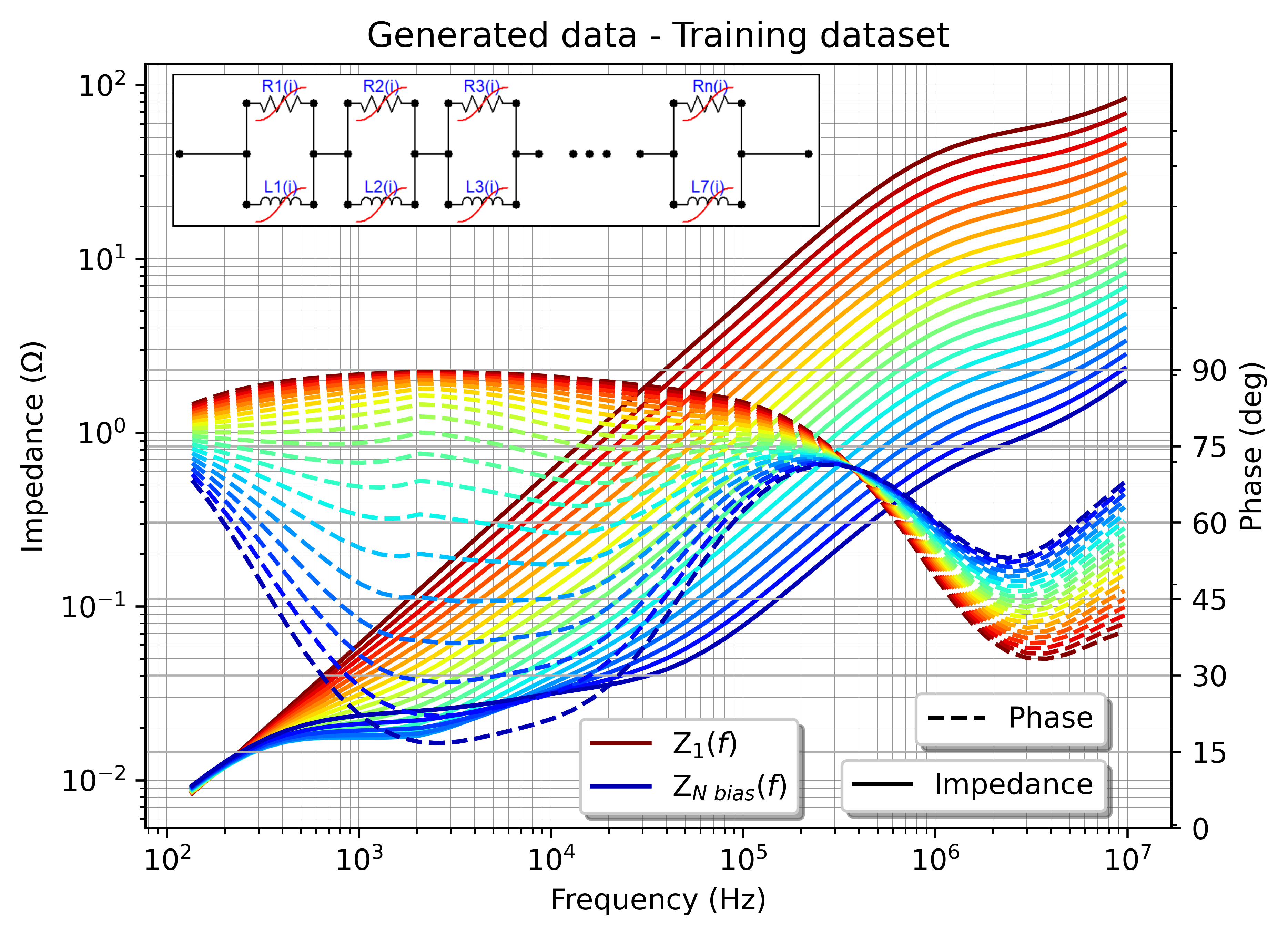} \\
    \end{tabularx}
        \caption{Impedance of a magnetic ring (a) measured \rd{(inset)} as a function of frequency for various \ra{DC current }biasings that controls operating point from \rd{linear range} ($I_\mathrm{DC}=0$~A, red curve) to the full saturation ($I_\mathrm{DC}=27$~A, blue curve), and (b) generated \ra{(random sample)} using lumped element model \rd{(inset)} with \rd{continuous transition between different curves (operating regimes)} \rd{and} randomly selected outermost characteristics (red and blue one) \ra{ and continuous transition between different curves)}. The generated data (b) will serve as a training set for the introduced neural network model, while the measurement ones (a) will be used to test the already trained NN model.}
        \label{fig:data}
\end{figure*}

To make our analysis realistic, we collect a set of measurements that will serve as a test set for the developed method. In Fig.~\ref{fig:data}(a), there is presented a family of impedances as a function of frequency $f$, taken for different \ra{DC-}bias current $I_\mathrm{DC}$ which controls the ring magnetization: from a linear range \ra{(see red curves in Fig.\mbox{\ref{fig:data}}a)} up to the ring saturation \ra{(blue curves in Fig.\mbox{\ref{fig:data}}a)}. The color of the measured impedance curves $Z^\mathrm{meas}$, that changes gradually from red, for $I_\mathrm{DC}=0$~A, to blue, for $I_\mathrm{DC}=27$~A, ($N_\mathrm{bias}=31$ different values in total) shows impedances for different \rd{biasings, combining this way multiple frequency characteristics controlled by the} \ra{DC-}bias current $I_\mathrm{DC}$ \ra{and frequency}: $Z^\mathrm{meas}=Z^\mathrm{meas}(I_\mathrm{DC},f)$.
Having families of characteristics, not just a single one, will be crucial when proving the model's capability to capture broad physics rather than just fitting a single feature. 
Methods for measuring magnetic rings over a wide range of frequencies and \ra{DC-bias current up to }saturation are not a trivial task (especially reaching the saturation limit is challenging) and are described in detail in Ref.~\cite{bib:MeasurementMR}.


Let us now introduce the analytical model of a ring in the form of LEEC impedance synthesized out of the series of $M=7$ L and R elements connected in parallel, as presented in Fig.~\ref{fig:data}(b), with the equivalent impedance in a form
\begin{equation}
Z^\mathrm{model}(s) = \sum_{i=1}^{M}\frac{sR_i}{s+\frac{R_i}{L_i}}, 
\label{eq:analytical_model}
\end{equation}
where $s=j 2\pi{}f$, and $j$ is the imaginary unit.
\ra{A series of parallel LR circuits was chosen as the best representation of the measurement data, whose impedance has a phase in the range of $0$-$90$ degrees (indicating it is resistive-inductive).} \ra{The number of $M=7$ elements was selected to ensure generality of the model and will be discussed in further sections.} The goal for the NN will be to learn the model defined in Eq.~\ref{eq:analytical_model} and predict its parameters, i.e., to find 
$\{L_i,R_i\}$ for a given frequency characteristics over the wide range of their possible behaviors. Therefore, to train the NN model and to force it to generalize over different behaviors, we have to synthesize the training dataset using the analytical model with random parameters. Generating data and training neural models on artificially synthesized datasets with controlled properties is currently a growing trend in machine learning~\cite{nikolenko2021}.

\begin{figure}[t]   
        \includegraphics[width=0.49\textwidth, trim={0cm -0cm 0 0.0cm}]{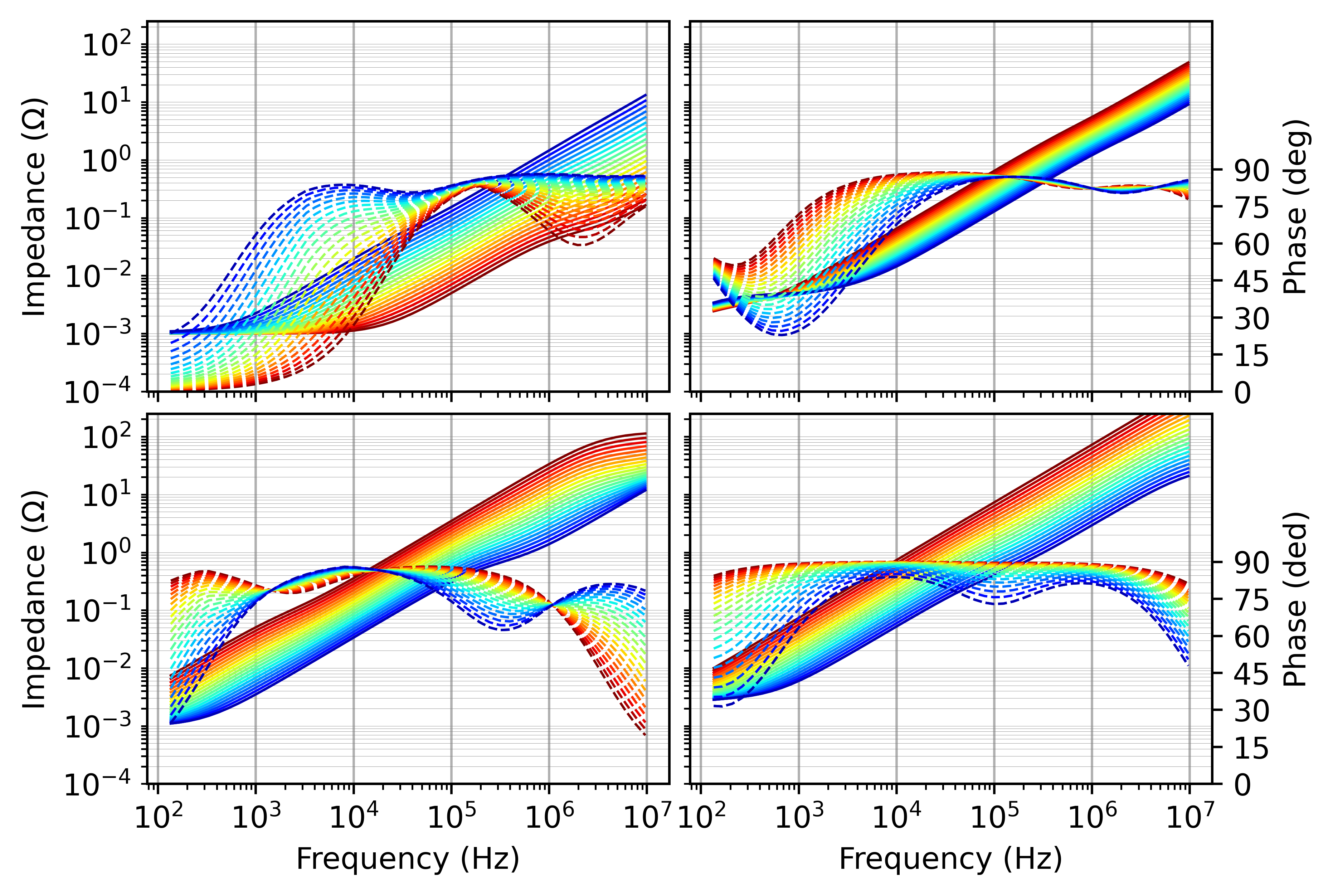} 
        \caption{\ra{Four representative families of characteristics $Z_j(I_\mathrm{DC},f)$, where $j=1\dots N_\mathrm{bias}$, from the synthesized training dataset. Solid line denotes impedance amplitude and dashed line denotes impedance phase as in Fig.\ref{fig:data}b}}
        \label{fig:extra}
\end{figure}

The synthesized dataset will be composed of families of $N_\mathrm{bias}$ frequency characteristics gradually transforming into each other, which mimics the situation of continuous transition in the ring material operating regime; in our measurements controlled by the \ra{DC-}bias current $I_\mathrm{DC}$. Each pair of the outermost characteristics $Z_1(f)$ \ra{(for zero DC-Bias)} and $Z_{N_\mathrm{bias}}(f)$ \ra{(for full saturation)} were generated from Eq.~\ref{eq:analytical_model} for two randomly selected sets of values (each of size $2M$): $R_i$ in the range from $10^{-3}$ to $10^4$~$\mathrm{\Omega}$, and $L_i$ in the range from $10^{-7}$ to $10^{-4}$~H. Assumed ranges correspond to \rd{marginal}\ra{boundary} values of these elements (equivalent resistance and inductance) during the measurements. 
Afterwards, for each pair of characteristics -- example set is presented in Fig.~\ref{fig:data}(b) with outermost ones marked in red and blue -- we calculate $N_\mathrm{bias}-2$ intermediate values at every frequency point
with equal distances for real and imaginary parts separately. This way we get $N_\mathrm{bias}$ characteristics that form continuous transitions (across their full $f$ range) from one outermost to another -- colors between red and blue in Fig.~\ref{fig:data}(b), in accordance to measurements with different \ra{DC-}bias currents. 
In total, we collected $N_\mathrm{train}=30~000$ multidimensional families, each containing $N_\mathrm{bias}=21$ curves gradually transforming into each other.
Examples of some other families of characteristics are presented in Fig.~\ref{fig:extra}.
These data will serve as the training set.

\section{Neural network model}

\ra{Machine learning (ML) is a rapidly expanding field that has begun to revolutionize many areas, including magnetic design in power electronics~\cite{magnet1,magnet2,corednn7,corernn}. ML is a branch of AI that focuses on developing algorithms and statistical models that enable computers to perform tasks without explicit instructions. Instead, they rely on patterns \textit{learned} directly from data. 
An \textit{autoencoder} (AE) is a type of NN used to learn efficient codings of data.
It is trained in an unsupervised manner  to ignore irrelevant “features” and learn to represent the data in a compressed form.
AEs are composed of two parts: \textit{encoder} processes the input data and compresses it into a smaller, dense representation, which captures the essence of the data, 
while \textit{decoder}, working in reverse of the encoder, takes the compressed representation and reconstructs the original input 
as closely as possible.}

\begin{figure*}[t]
\includegraphics[width=\textwidth]{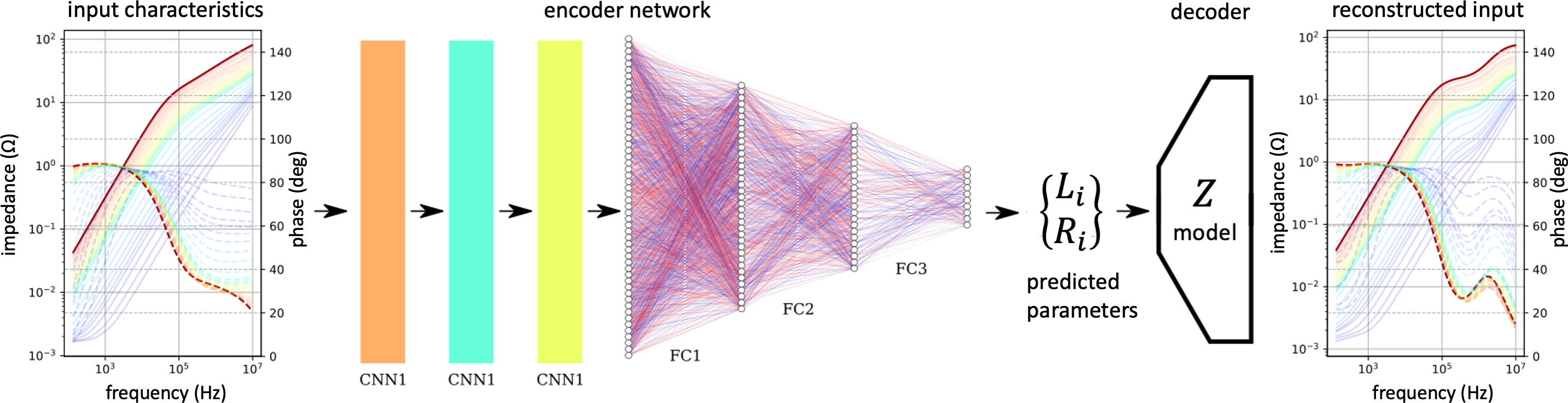}
\caption{Neural network basic model with an autoencoder structure, which is of fundamental type and used in various fields of machine learning or image processing. The encoder is made of three convolutional layers (CNN) and three fully connected (FC) layers. The CNN layers are used as feature extractors, while the FC layers are used for parameters ($\{L_i,R_i\}$) predicting based on previously found representations. The decoder ($Z$) just implements an analytical formula for the impedance of the LEEC model. The goal of the network is to reconstruct the input characteristics at the output.}
\label{fig:model_sieci_0}
\end{figure*}

After defining the generated training and measured test datasets, let us define the NN responsible for finding the model parameters. The network will take a form of AE as illustrated in Fig.~\ref{fig:model_sieci_0}, and trained in an unsupervised way. The role of the encoder is to predict the analytical model parameters $\{L_i,R_i\}$ (their values define the \textit{latent space} of our AE), while the decoder will just calculate the frequency characteristic directly using Eq.~\ref{eq:analytical_model} with parameters (latent space) found by the encoder. 

\rd{The goal for the NN is to reconstruct input characteristics as best as possible, with the smallest reconstruction error called \textit{loss} -- this means that in our training we do not need any labels and only frequency characteristics to be reconstructed which makes our training scheme fully unsupervised.} \ra{The goal for the NN is to reconstruct input characteristics as accurately as possible, minimizing the reconstruction error, defined by the \textit{loss} function. This means that in our training process, we don't require any true predictions (\textit{labels}); only the frequency characteristics need to be reconstructed. This makes our training scheme fully unsupervised.}

The standard solution would also be to use an NN as the decoder, but this approach makes less sense because we need the decoder only to train the encoder, which in turn, is essential for us (it predicts the analytical model parameters). Therefore, we successfully replaced the NN decoder with the analytical model itself (here defined by Eq.~\ref{eq:analytical_model} for equivalent LEEC impedance) implemented as part of the NN (without trainable parameters, but using the model parameters ($\{L_i,R_i\}$) predicted by the encoder) to be able to use the standard \textit{backpropagation} of loss mechanism during the training.

Frequency characteristics are organized on a 1D evenly spaced grid along multiple decades on the log scale ($50$ points from $135$~Hz to $10^7$~Hz).
To build an effective neural representation, we used convolutional neural network (CNN) layers as feature extractor parts, which is a common way to process data organized on regular grids~\cite{zeiler2014visualizing,Yu2016dilated}. Then, we applied several fully connected (FC) layers serving as parameter predictors, which are based on previously found representations (by CNN).
The detailed structure of the encoder network is presented in Fig.~\ref{fig:model_sieci_0}: three CNN feature extractor layers are followed by three FC layers with $2M$ \ra{(see. Eq.\ref{eq:analytical_model})} outputs in the last layer that predict the parameters.
We select CNN as a feature extractor (backbone) part, which is common practice in signal or image processing~\cite{Alam2020,Chai2021}; however, in principle, any other type of network could serve as backbone network at this stage, depending on applications -- such as recurrent neural networks or nowadays very popular transformer networks~\cite{attentionpaper} with a self-attention mechanism incorporated.

To train a NN model, we have to define a loss function $\mathscr{L}$, which in our case will be just the decoder (reconstruction) loss $\mathscr{L}=\mathscr{L}_\mathrm{dec}$, i.e. the mean squared difference between input and output impedance, taken in logarithmic scale for real and imaginary part separately: 
\begin{equation}
\mathscr{L}_\mathrm{dec}=\mathrm{MSLE}\left(\Re(Z),\Re(\hat{Z})\right)
+\mathrm{MSLE}\left(\Im(Z),\Im(\hat{Z})\right),
\label{eq:loss_decoder}
\end{equation}
with $\hat{Z}$ meaning reconstructed characteristic (decoder output), and
mean squared logarithmic error defined as
\begin{equation}
\mathrm{MSLE}\left(Z,\hat{Z}\right) = \frac{1}{N}\sum_{j=1}^{N}
\left(\log(Z_j)-\log(\hat{Z}_j)\right)^2.
\label{eq:msle}
\end{equation}
The AE is trained over the whole training set and
the sum in Eq.~\ref{eq:loss_decoder} goes over $N=N_\mathrm{train}\times N_\mathrm{bias}$ frequency characteristics $Z$ in the training set. However, in practice, during a single training step of the stochastic gradient descent (SGD) method, only a smaller random portion (mini-batch) is used, here $N_\mathrm{batch}=1024$. Training typically lasted $100$-$500$ epochs up to the loss saturation. To validate the NN model performance after the training, we also define the relative error of the fitting (FE):
\begin{equation}
\mathrm{FE}=\sqrt{\mathrm{RE}\!\left(\Re(Z),\Re(\hat{Z})\right)+\mathrm{RE}\!\left(\Im(Z),\Im(\hat{Z})\right)},
\label{eq:fit_error}
\end{equation}
with
\begin{equation}
\mathrm{RE}\left(Z,\hat{Z}\right) = \frac{1}{N}\sum_{j=1}^{N}\left(\frac{\log(Z_j)-\log(\hat{Z}_j)}{\log(Z_j)}\right)^2.
\label{eq:fit_error_1}
\end{equation}

\section{Basic model results}

Let us now analyze the NN model performance. To verify the model's ability to generalize over the broad variety of characteristics we estimate the validation error using Eq.~\ref{eq:fit_error} but calculated after the training on some additional validation set generated in the same way as the training one. Table~\ref{tab:val_performance} proves that the model is capable of predicting parameters for a variety of characteristics (as example one presented in Fig.~\ref{fig:data}(b)), which is a crucial property, a necessary condition that must be met in order to have a chance of predicting parameters for unknown measured characteristic.
\begin{table}[b]
\centering
\begin{tabular}{|ccc|}
\hline
\multicolumn{3}{|c|}{models validation error} \\ \hline
\multicolumn{1}{|c|}{basic} & \multicolumn{1}{c|}{Siamese} & modified Siamese \\ \hline
\multicolumn{1}{|c|}{8.24\%} & \multicolumn{1}{c|}{5.29\%} & 4.54\% \\ \hline
\hline
\multicolumn{3}{|c|}{models test error} \\ \hline
\multicolumn{1}{|c|}{basic} & \multicolumn{1}{c|}{Siamese} & modified Siamese \\ \hline
\multicolumn{1}{|c|}{12.54\%} & \multicolumn{1}{c|}{7.22\%} & 5.18\% \\ \hline
\end{tabular}\vspace{1mm}
\caption{Relative fitting error for different modifications of the NN model calculated using validation synthetic set and measured test set.}
\label{tab:val_performance}
\end{table}
The result presented in 1st column of Table~\ref{tab:val_performance} shows that the basic version of the NN model (depicted in Fig.~\ref{fig:model_sieci_0}) generalizes quite well with the fitting error slightly above $8\%$.

\begin{figure*}[t]
    \centering

    \begin{tabularx}{\textwidth}{ p{-2.6cm} l  p{-2.6cm}  l }
    (a) &   & (b) &  \\ 
         & \includegraphics[width=0.46\textwidth, trim={1.1cm 0 0 0.1cm}]{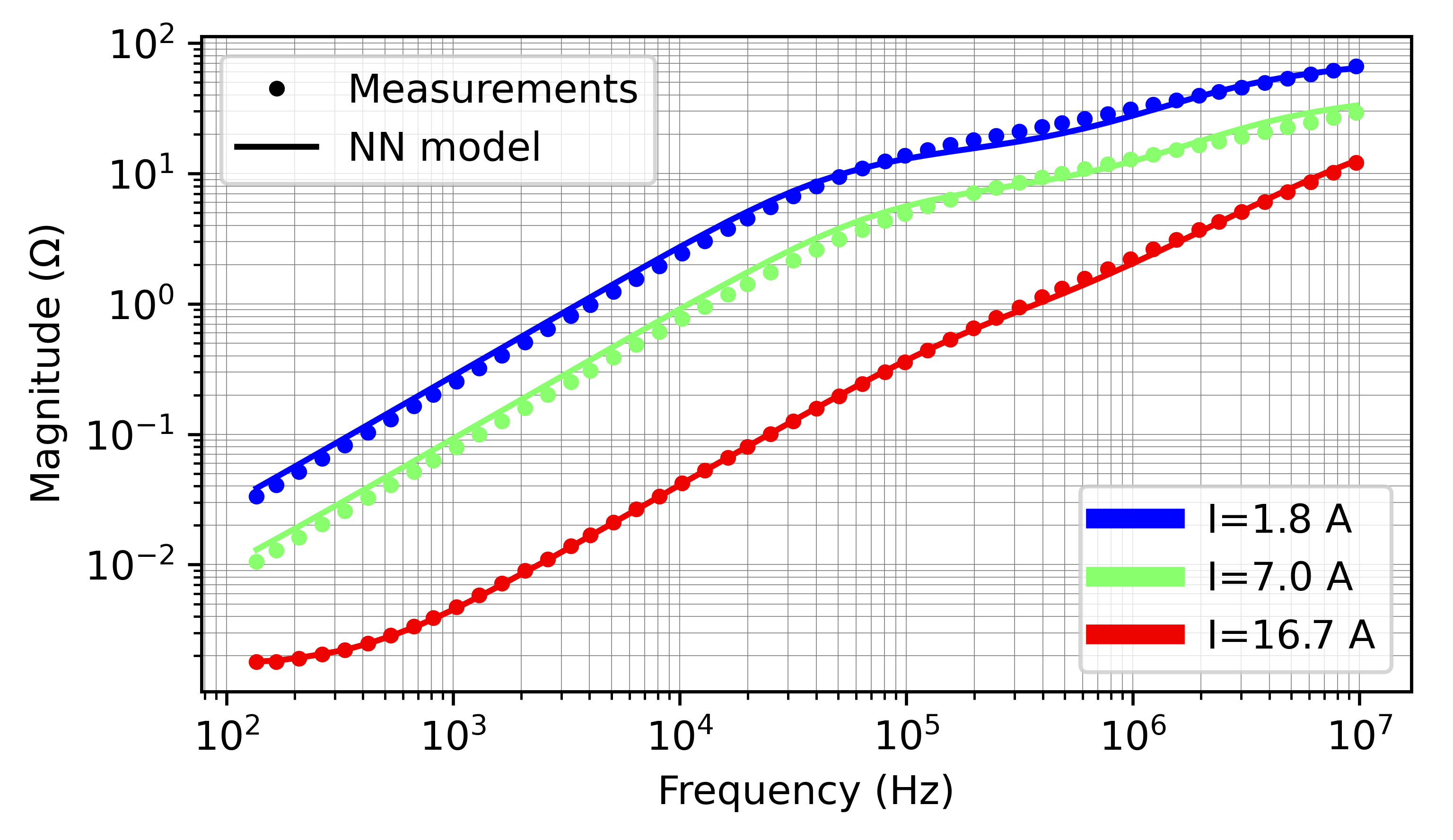} & &\includegraphics[width=0.46\textwidth, trim={0.9cm 0 0 0.1cm}]{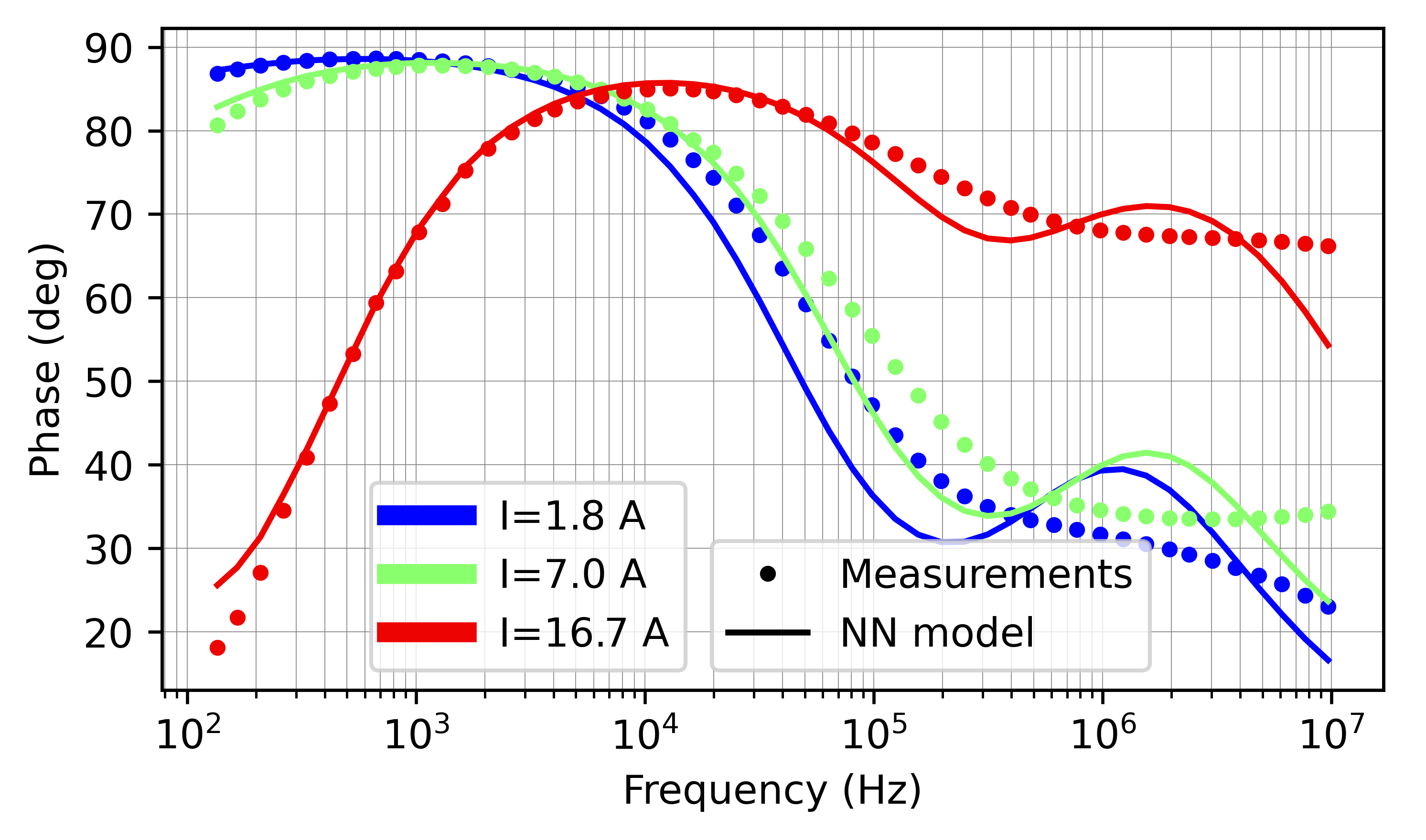} \\
          (c) &   & (d) &  \\ 
         & \includegraphics[width=0.46\textwidth, trim={1.1cm 0 0 0.1cm}]{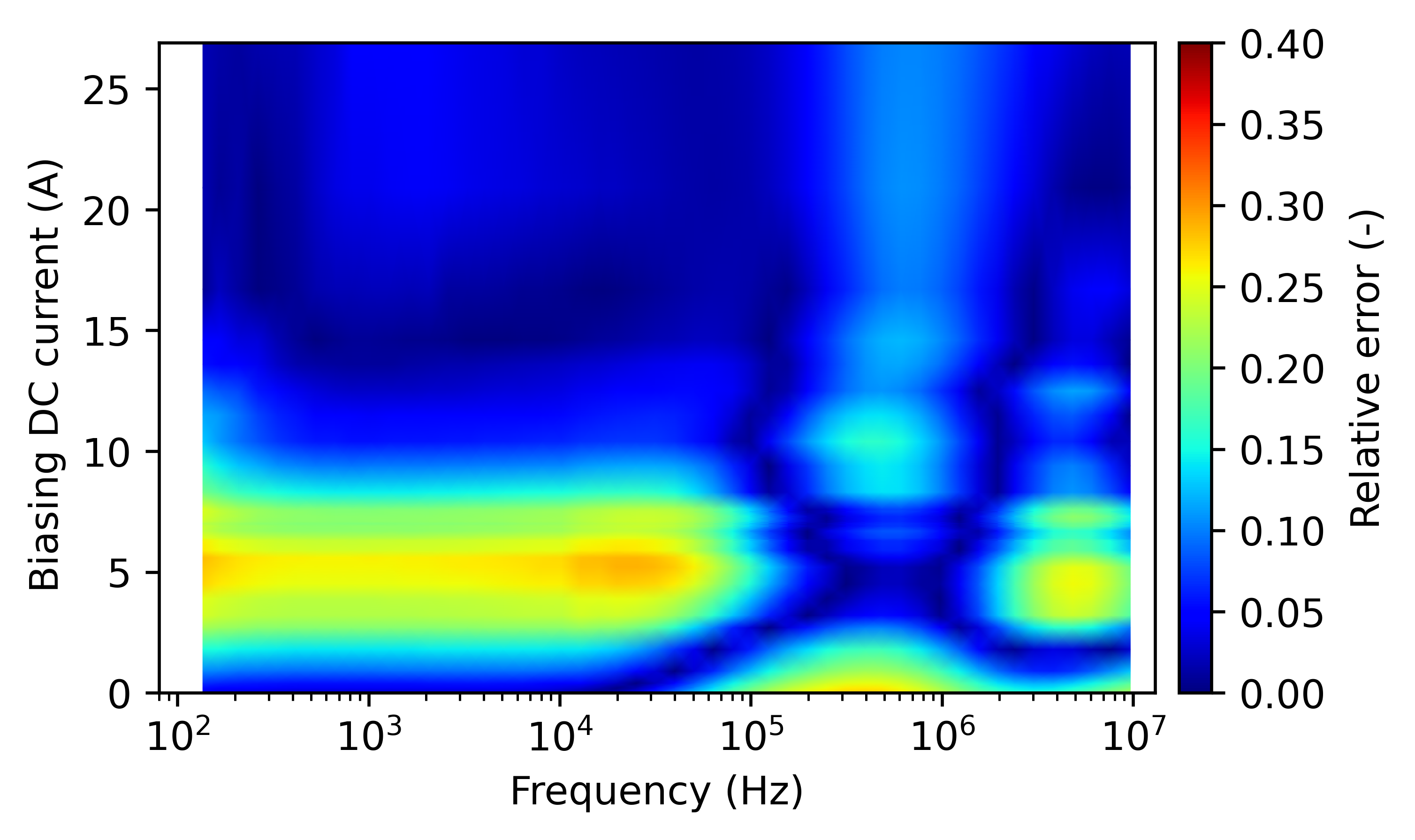} & &\includegraphics[width=0.46\textwidth, trim={0.9cm 0 0 0.1cm}]{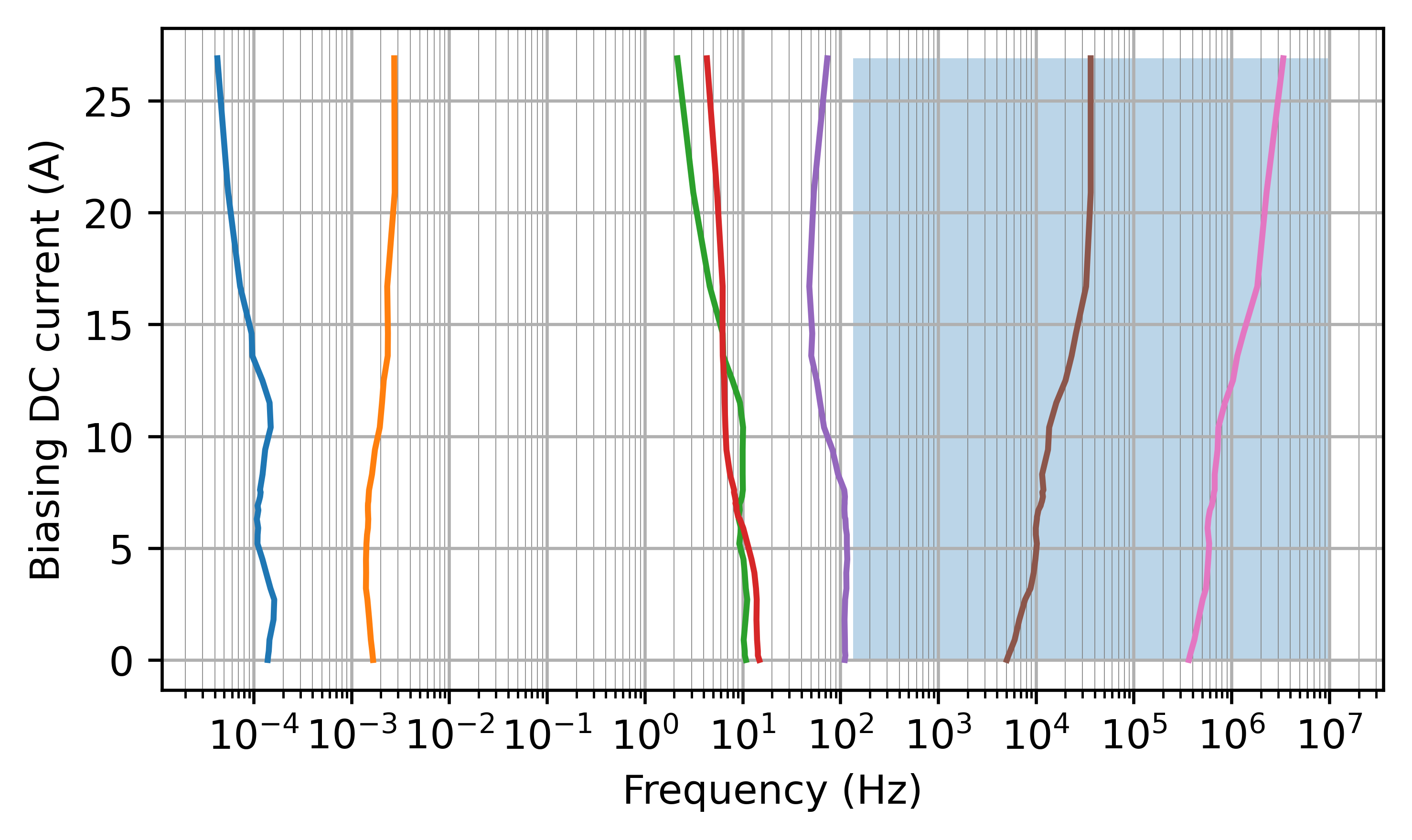} 
         
    \end{tabularx}
    
    \label{fig:first}
        \caption{
        Results for testing the basic neural network model on the measured ring data: (a,b) examples of fitting the analytical model (solid curves), with parameters predicted by the NN model, versus measurements (dots) for the selected \ra{DC-}bias currents $I_\mathrm{DC}=\rd{\{0.9, 6.9, 12.5\}}\ra{\{1.8, 7.0, 16.7\}~\mathrm{A}}$, (c) relative fitting error for the various \ra{DC-}bias currents $I_\mathrm{DC}$ plotted versus frequency, and (d) characteristic frequencies \ra{$f_i=2\pi R_i/L_i$ of the predicted NN model parameters (each line represent $f_i$ for one LR pair in the LEEC series)} \rd{of the predicted parameters (each representing LR pair in the LEEC series)} versus \ra{DC-}bias current $I_\mathrm{DC}$.\ra{The blue color marks the frequency range where the measurements data are located. }. 
        }
        \label{fig:result_for_model_0}
\end{figure*}

Let us now perform tests and apply the model to \rd{realistic,} measured data as presented in
Fig.~\ref{fig:data}(a). Comparison of characteristics for the analytical model (defined in Eq.~\ref{eq:analytical_model}, with parameters predicted via the NN model) and the measured one, for selected $I_\mathrm{DC}$ values, are presented in Fig.~\ref{fig:result_for_model_0}(a,b) and shows quite good fitting performance over the broad range of \ra{DC-}bias currents. It is worth recalling here that the measurement data was not a part of the training dataset. Moreover, we calculated the relative fitting error systematically, using Eq.~\ref{eq:fit_error} but now on the measured data, for every \ra{DC-}bias current over the frequency range as presented in Fig.~\ref{fig:result_for_model_0}(c). In general, we observe that the error does not exceed $30\%$, which is quite a resonable result for fitting the model over so wide range\rd{s} of frequencies, impedance amplitudes, and \ra{DC-}bias currents $I_\mathrm{DC}$. On average it is even smaller and equals $12.54\%$.

\section{Siamese neural network model}
\begin{figure*}[h]
\centering
\includegraphics[width=0.9\textwidth]{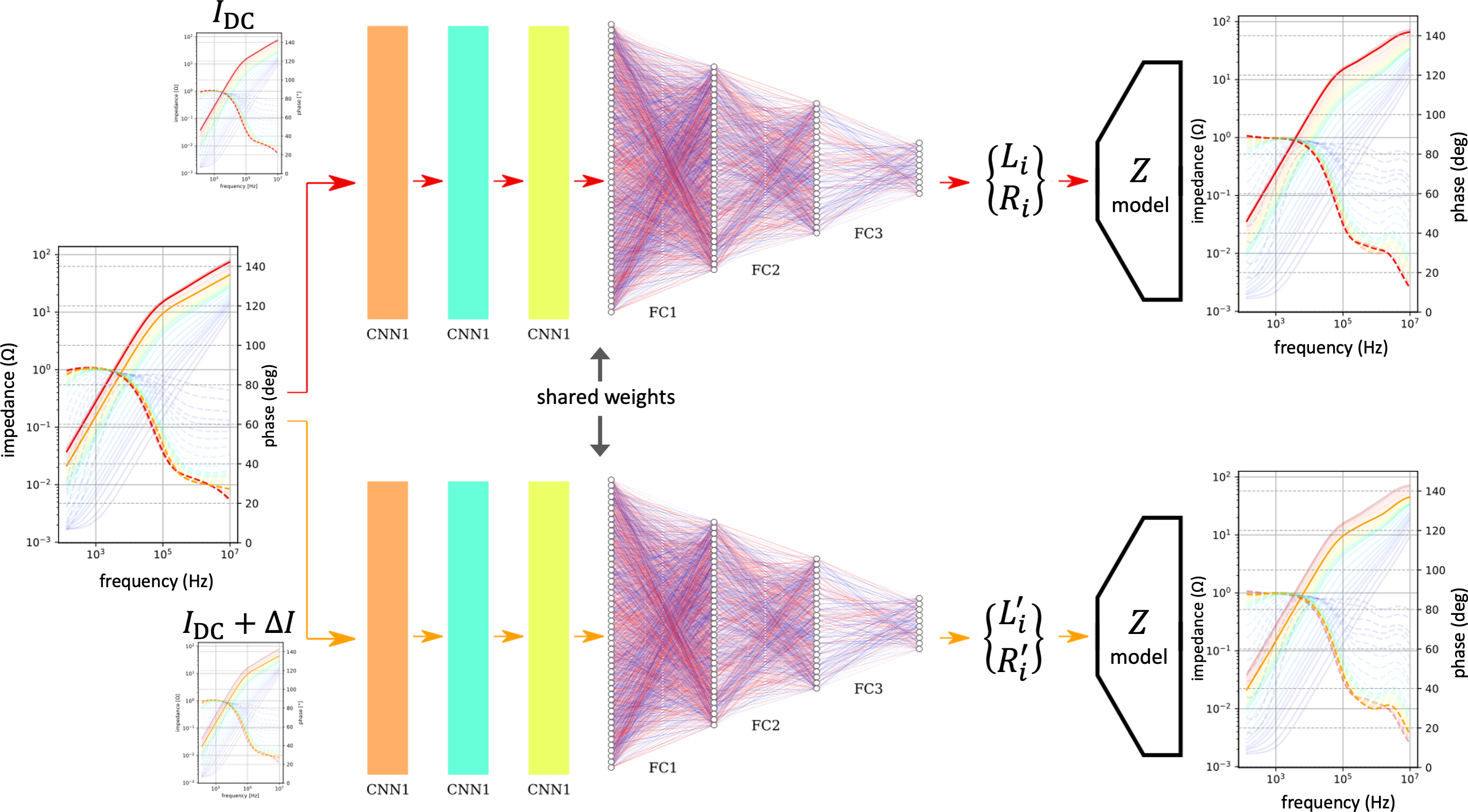}
\caption{Siamese neural network model composed of two copies of the basic model that share the weights and process two neighbor characteristics (that differ in \ra{DC-}bias current by $\Delta I$) at the same time enabling to train the model to predict similar (or continuously changing) parameters for similar characteristics.}
\label{fig:model_sieci_2}
\end{figure*}

The results obtained for the basic NN model are acceptable but have one substantial drawback.
If we look at Fig.~\ref{fig:result_for_model_0}(d) which presents values of characteristic frequencies defined as $\omega_i = R_i/L_i$ for each of LR pairs in a series (of size $M=7$) for subsequent $I_\mathrm{DC}$  currents, we observe that the values are irregular and quite independent from each other which seems to be unphysical. In order to restore the more realistic behavior which is more likely to be continuous (we do not expect to have any phase transitions in the model here), the \rd{single}\ra{standard} NN model was extended to the Siamese model shown in Fig.~\ref{fig:model_sieci_2}.
The idea behind the Siamese NNs, popular in many ML areas, such as face recognition~\cite{FaceNet}, is to have two (or more) copies of a network that works in parallel and process different data samples at the same time. 
\ra{Two Siamese copies use (share) the same NN parameters (weights) while working in tandem (see Fig.~\ref{fig:model_sieci_2}) on two different input samples to compute two separate predictions. Throughout the training, the weights are effectively updated twice by applying distinct loss terms for each branch, along with an additional term that just compares the two predictions.}
Processing multiple samples at once enables to regularize of the model behavior with respect to the similarity between data samples by adding this extra term to the loss function. In our case we can process neighbor characteristics, i.e. for $I_\mathrm{DC}$ and $I_\mathrm{DC}+\Delta I$ -- see Fig.~\ref{fig:model_sieci_2}, and enforce parameters predicted for that data points to change continuously, by including additional loss terms in a form:
\begin{equation}
l_\mathrm{cont}(L) = \frac{1}{N_\mathrm{bias}-1}\sum_{k=1}^{N_\mathrm{bias}-1}\log^2\!\left(\frac{L_{k+1}}{L_{k}}\right).
\label{eq:loss_cont}
\end{equation}
The sum in Eq.~\ref{eq:loss_cont} goes for every $N_\mathrm{bias}-1$ neighboring pair in a given characteristics family. It penalizes too high a change of the parameters between neighbors.
The total continuity loss should include $M$ parameters and all $N_\mathrm{train}$ training families:
\begin{equation}
\mathscr{L}_\mathrm{cont}(L) = \frac{1}{M N_\mathrm{train}}
\sum_{i=1}^{M}\sum_{j=1}^{N_\mathrm{train}}
l_\mathrm{cont}(L_{i,j}).
\label{eq:loss_cont_tot}
\end{equation}
In the same way, we define $\mathscr{L}_\mathrm{cont}(R)$.
Moreover, we found that in our case it is beneficial to add yet another term to the loss function:
\begin{equation}
l_\mathrm{mono}(L)=
    \left\{\begin{matrix}
        0  \text{\hspace{3mm}if\hspace{3mm}} \frac{1}{N_\mathrm{bias}-1}\sum_{k=1}^{N_\mathrm{bias}-1}(L_{k+1}-L_{k}) < 0, \\
        \\
        \frac{1}{N_\mathrm{bias}-1}\sum_{k=1}^{N_\mathrm{bias}-1}(L_{k+1}-L_{k}) \text{\hspace{3mm}otherwise,}
    \end{matrix}\right. 
    \label{eq:loss_L}
\end{equation}
in order to maintain a monotonic decrease in the parameter values with the \ra{DC-}bias current $I_\mathrm{DC}$, which is justified by the ring material saturation.
The total monotonicity loss $\mathscr{L}_\mathrm{mono}(L)$ is defined in the same way as in Eq.~\ref{eq:loss_cont_tot}. 
All in all, the loss function for the Siamese NN model is as follows:
\begin{align}
\mathscr{L}_\mathrm{Siam} = 
\mathscr{L}_\mathrm{dec} 
+& \alpha_1\mathscr{L}_\mathrm{cont}(L) 
+ \alpha_2\mathscr{L}_\mathrm{cont}(R)+\nonumber\\ 
+& \alpha_3\mathscr{L}_\mathrm{mono}(L)  
+ \alpha_4\mathscr{L}_\mathrm{mono}(R),
\label{eq:loss_7}
\end{align}
with the training hyperparameters $\alpha_i$ that control the impact of the loss components responsible for keeping the continuity of the parameters under $I_\mathrm{DC}$ changes.

\begin{figure*}[h]
    \centering
            \begin{tabularx}{\textwidth}{ p{-2.6cm} l  p{-2.6cm}  l }
    (a) &   & (b) &  \\ 
         & \includegraphics[width=0.46\textwidth, trim={1.1cm 0 0 0}]{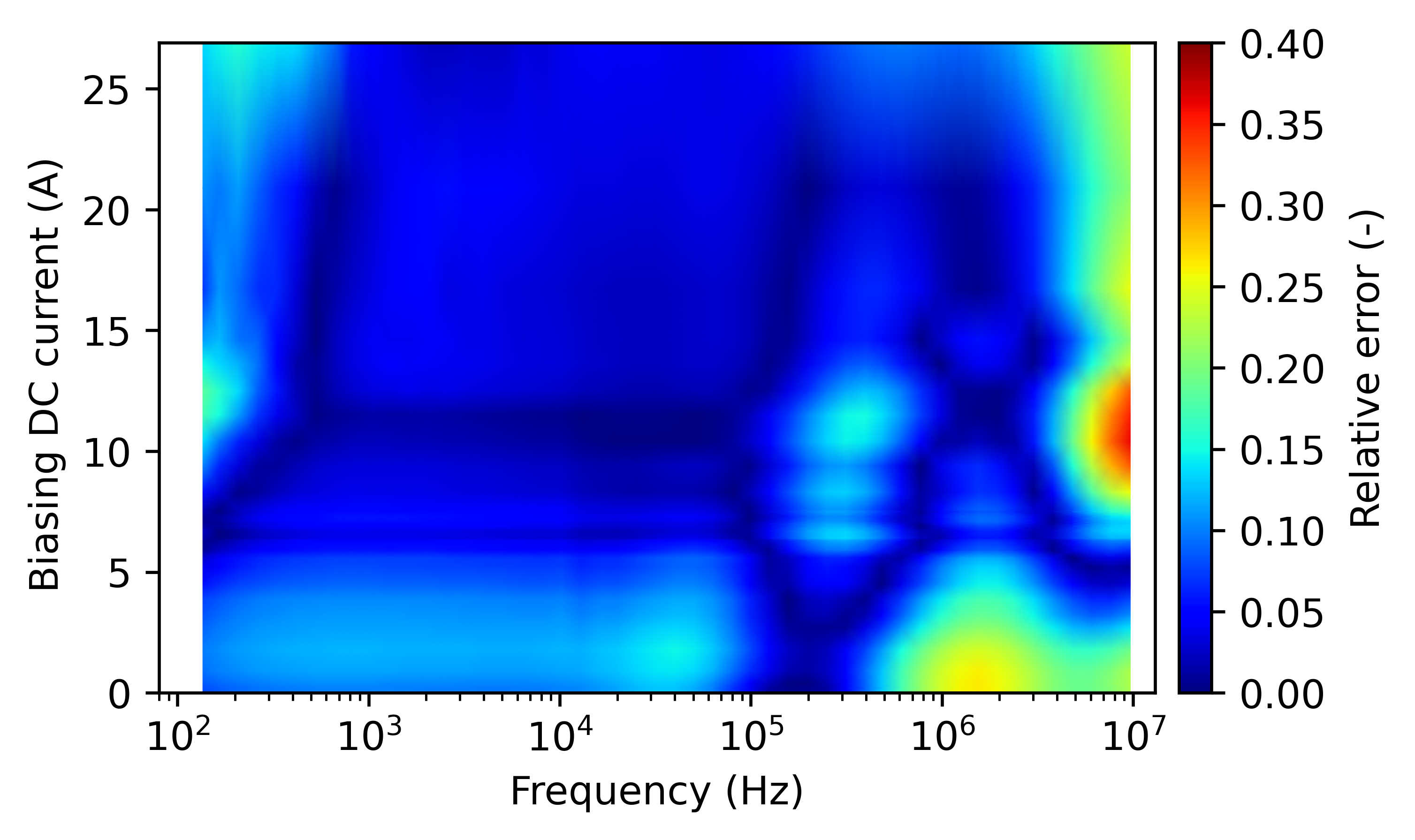} & &\includegraphics[width=0.46\textwidth, trim={0.9cm 0 0 0}]{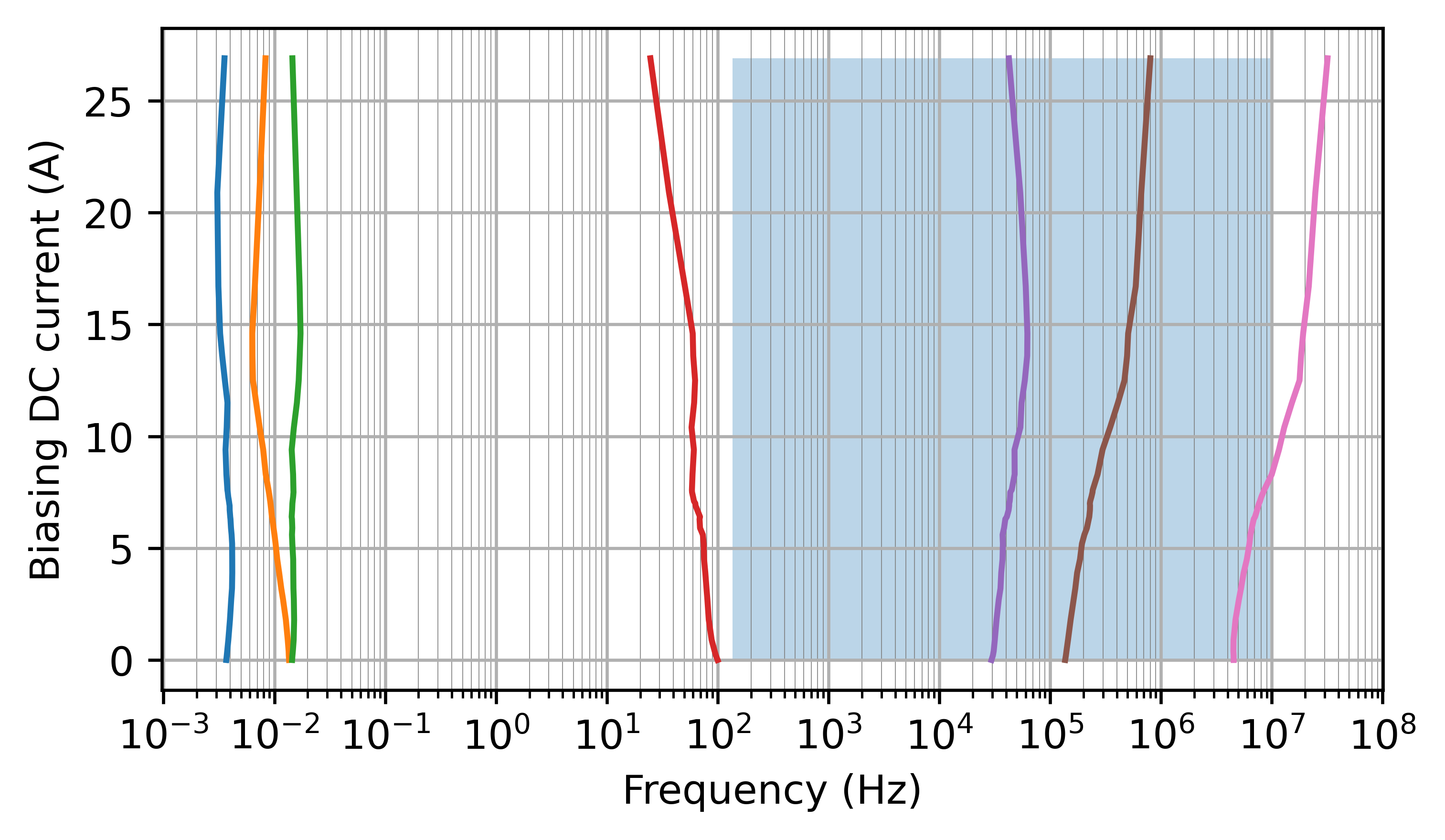} \\
          (c) &   & (d) &  \\ 
         & \includegraphics[width=0.46\textwidth, trim={1.1cm 0 0 0}]{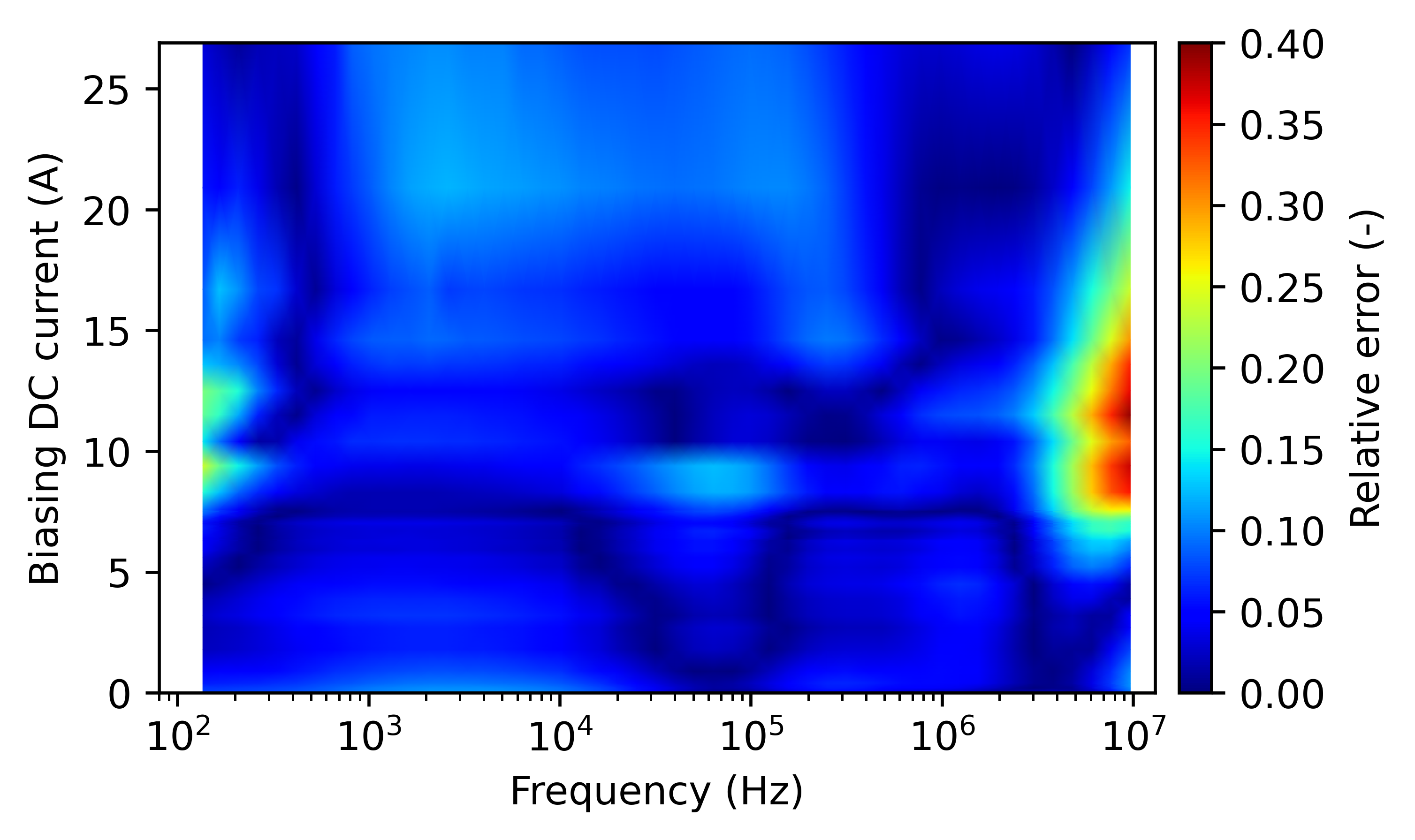} & &\includegraphics[width=0.46\textwidth, trim={0.9cm 0 0 0}]{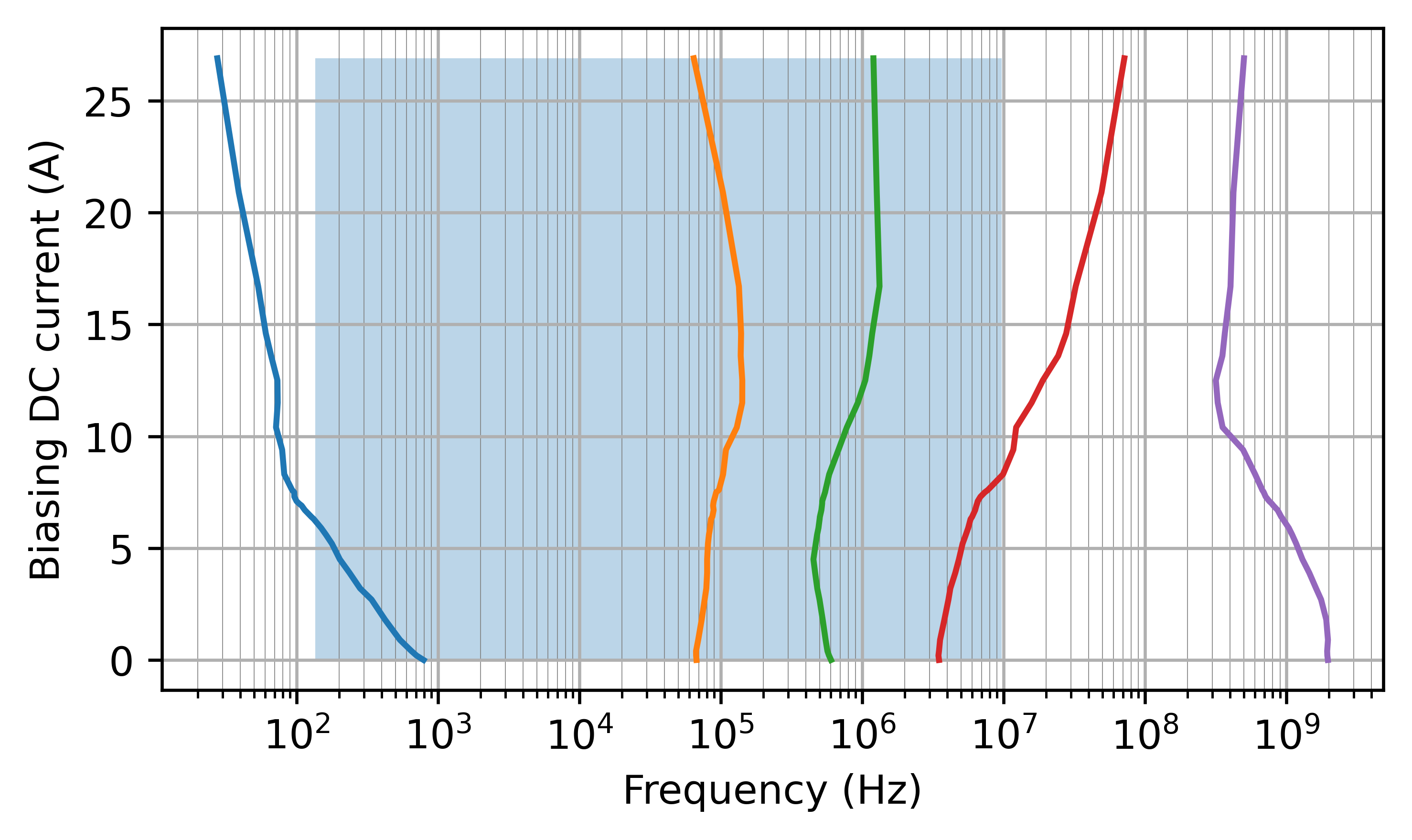} 
    \end{tabularx}
        \caption{Same as in Fig.~\ref{fig:result_for_model_0} but results for (a,b) Siamese neural network, and (c,d) modified Siamese model.}
        \label{fig:result_for_model_7_and_5}
\end{figure*}

If we look at the results for the Siamese NN model presented in Fig.~\ref{fig:result_for_model_7_and_5}(a,b), where we force the model to predict similar parameters for similar characteristics,  we can observe that now characteristic frequencies $\omega_i=R_i/L_i$ change with the \ra{DC-}bias current $I_\mathrm{DC}$ continuously and in some case monotonically -- see Fig.~\ref{fig:result_for_model_7_and_5}(b). Moreover,
a surprising but also very desirable property is that forcing the continuity for the change of the parameters $\{L_i,R_i\}$ with $I_\mathrm{DC}$ at the same time has made fitting of the analytical model better than before -- in Fig.~\ref{fig:result_for_model_7_and_5}(a) the relative fitting error typically does not exceed $15\%$ with the average value over frequencies and currents equals $7.22\%$.   

\section{Modified Siamese model}
 
Fig.~\ref{fig:result_for_model_7_and_5}(b)
presents yet another feature of the Siamese model (which was also partially present in the basic NN model) that it has a strong tendency to select features-parameters of the predicted analytical model (Eq.~\ref{eq:analytical_model}). It manifests in the fact that some groups of parameters, i.e. LR pairs in a series, are irrelevant -- their characteristic frequencies, here $\omega_1$, $\omega_2$, $\omega_3$, lie outside the ring working frequencies, which means that they can be safely skipped in the model. This motivates us to reduce the dimensionality of the model by assuming $M=5$. Moreover, we observed that enforcing a more uniform positioning of the characteristic frequencies in the frequency domain over which the model operates can be beneficial in terms of further reducing the model size.
Therefore, we added another component to the loss function of the form:
\begin{equation}
\mathscr{L}_{\omega} =  \frac{1}{NM} \sum_{j=1}^{N}\sum_{i=1}^{M}\left(\frac{R_i}{L_i}-\Omega_i\right)^2,
\label{eq:loss_omega}
\end{equation}
with $N=N_\mathrm{train}\times N_\mathrm{bias}$, $M=5$, and the constant frequency nodes centered at
positions $\Omega_i=\{10^3,10^{4.75},10^{6.5},10^{8.25},10^{10}\}$~Hz. \ra{The values of \(\Omega_i\) have been chosen to evenly cover the frequency range of the measurement data. Moreover, including an extra point above this range further improved the fitting.}
Now the loss function for the modified Siamese NN model is the following:
\begin{equation}
\mathscr{L}_\mathrm{Siam'} = \mathscr{L}_\mathrm{Siam} + \alpha_5 \mathscr{L}_\mathrm{\omega}.
\label{eq:loss_5}
\end{equation}
The results in Fig.~\ref{fig:result_for_model_7_and_5}(c,d) clearly show that now the positioning of the characteristic frequencies is more homogeneous and all of $\omega_i$ occupies ring working frequency range. In addition, the relative fitting error obtained here is less than $10\%$ in most cases and on average is the smallest among all three NN models, reaching $5.18\%$.

\section{Discussion and Conclusions}
In the first approach to the problem, we defined a network that predicts the parameters directly (containing the encoder part only) and compares them with the parameters used to generate a given characteristic. Such training was strictly supervised by using the encoder loss, defined as\ra{:}
\begin{equation}
\mathscr{L}_\mathrm{enc} = \frac{1}{2NM}\sum_{j=1}^{N}\sum_{y_i\in\{L_i,R_i\}}^{2M}
\log^2\left(\frac{y_{ij}}{\hat{y}_{ij}}\right).
\label{eq:loss_encoder}
\end{equation}
The sum in Eq.~\ref{eq:loss_encoder} goes over the training samples and parameters: $\hat{y}_{ij}$ means the $i$-th parameter ($L_i$ or $R_i$) predicted for $j$-th sample, $Z_j$, whereas $y_{ij}$ represents ground truth parameters -- we already know them from the generation process.
However, this approach simply failed because it was not possible to effectively train a neural model that would generalize beyond a limited set of characteristics being some narrow part of the training data. 
Much better results came from the unsupervised training where the AE model is simply given a variety of characteristics and has to attempt to reconstruct them. The model proposed in our scheme is able to generalize well, which is shown in Table~\ref{tab:val_performance} where the validation error reaches a few percent for the training set with a huge variety of characteristics. Consequently, the NN model is able to correctly fit parameters for previously unseen measured test data. At the same time, it is worth noting that the diversity of the training set (see Fig.\ref{fig:extra}) fulfilled a kind of regularization, ensuring that the NN does not overfit any particular behavior.

\ra{In the proposed approach, the decoder is needed only to estimate the loss function ($\mathcal{L}_\mathrm{dec}$), that is minimized  during the training. The approach where a large component of NN model (in our case, the decoder) is used only to estimate the error of the predictive part (encoder) in ML is not unusual. It is characteristic, for example, for generative adversarial networks (GANs)\cite{gan}, where the discriminator is trained to estimate the error of the generator, or in neural style transfer models~\cite{styletransfer}, where the pretrained NN determines so-called style or content losses.}

\ra{The proposed scheme for fitting parameters of lumped model is general and in principle can be applied to any analytical model. The decoder introduces analytical formula which parameters are predicted by the encoder. The only requirement is that the decoder formula, which will pass the fitting error when the encoder is trained by backpropagation, should be differentiable.}

In the proposed framework the materials are represented by multidimensional features, in our case as normalized impedance data in the form of real and imaginary parts calculated from the modulus and phase values of the data generated. The way the NN is trained is designed so that we can naturally enforce continuous (or any other physical) behavior on the entire family of measurement characteristics simultaneously.

\rd{We also observe the important effect that imposing continuity of the predicted parameters as a function of the \ra{DC-}bias current using the Siamese network, i.e., forcing the current-dependent relations to be physical, causes the fit to individual characteristics in the family to improve significantly. }

\ra{In classical methods for finding equivalent circuit elements, there is difficulty in capturing the correlation between individual $Z(f)$ characteristics for different DC-bias currents \cite{bib:NMMcircuitSimulation}, and they are usually only able to match individual characteristics \cite{szewczyk2014identification,gustavsen1999rational}. This results in discontinuous parameters (here, the ladder elements L, R) as a function of the DC current. The Siamese network proposed here solves this problem.
Vector fitting, as a standard method for finding transfer function parameters, not only has the problem of ensuring continuous parameters as a function of DC current but also cannot guarantee that the identified transfer function is a positive real function\cite{de2008single}. Using neural networks allows for the imposition of arbitrary constraints and boundary conditions, including enforcing the physical behavior of the fitted parameters.}

\ra{The proposed method exhibits a robust tendency to parameter selection. Although the initially suggested complexity of analytical model was $M=7$, the NN network identified some components as having negligible contributions (as shown in Fig. 5(b)), effectively reducing the complexity $M\rightarrow{}4$. Consequently, in the subsequent experiment (Fig. 5(d)), NN successfully predicted a simplified analytical model with $M=5$ while maintaining a comparably low error.}

\ra{One way to further improve the presented model is to eliminate the arbitrarily determined grid of \(\Omega_i\) (Eq.~\ref{eq:loss_omega}) by adding a term \ra{in loss function} that penalizes the characteristic frequencies of the individual LR ladder elements coming too close to each other. This will result in a more uniform distribution of these frequencies without imposing their positions from above.}

\section{Methods}
Data generation was performed with the usage of
Pytorch library~\cite{paszke2019pytorch} on GPU \ra{(RTX A6000 48GB GDDR6)}, which significantly speedups the generation process from a few hours (when using CPU\ra{ - 64 cores 128 threads Ryzen Threadripper 3990X}) to a dozen seconds (GPU). NN model (encoder and decoder part) was implemented using Pytorch with an automatic differentiation engine (autograd) used to
track the gradient of the loss with respect to the model parameters during the backpropagation stage. The training of the single model on the GPU Nvidia \ra{RTX} A6000 took about 1 hour.

Measurements were taken for NANOPERM nanocrystalline rings from Magnetec GmbH labeled M-676~\cite{magnetec_datasheet_1}, for different magnetization currents $I_\mathrm{DC}$ up to ring saturation using the Keysight Technologies Impedance Analyzer E4990A-030 to measure the impedance frequency characteristic and Keysight Technologies DC Power Supply N8731A to supply the DC-bias current $I_\mathrm{DC}$.

\section{Acknowledgments}

The authors would like to acknowledge the financial support of Project No. 2019/34/E/ST7/00187 granted by National Science Centre, Poland (NCN).



\bibliographystyle{IEEEtran}
\bibliography{main}

\begin{thebibliography}{10}
\providecommand{\url}[1]{#1}
\csname url@samestyle\endcsname
\providecommand{\newblock}{\relax}
\providecommand{\bibinfo}[2]{#2}
\providecommand{\BIBentrySTDinterwordspacing}{\spaceskip=0pt\relax}
\providecommand{\BIBentryALTinterwordstretchfactor}{4}
\providecommand{\BIBentryALTinterwordspacing}{\spaceskip=\fontdimen2\font plus
\BIBentryALTinterwordstretchfactor\fontdimen3\font minus
  \fontdimen4\font\relax}
\providecommand{\BIBforeignlanguage}[2]{{%
\expandafter\ifx\csname l@#1\endcsname\relax
\typeout{** WARNING: IEEEtran.bst: No hyphenation pattern has been}%
\typeout{** loaded for the language `#1'. Using the pattern for}%
\typeout{** the default language instead.}%
\else
\language=\csname l@#1\endcsname
\fi
#2}}
\providecommand{\BIBdecl}{\relax}
\BIBdecl

\bibitem{li2023magnet_1}
H.~Li, D.~Serrano, T.~Guillod, S.~Wang, E.~Dogariu, A.~Nadler, M.~Luo,
  V.~Bansal, N.~K. Jha, Y.~Chen \emph{et~al.}, ``How magnet: Machine learning
  framework for modeling power magnetic material characteristics,'' \emph{IEEE
  Transactions on Power Electronics}, 2023.

\bibitem{serrano2023magnet}
D.~Serrano, H.~Li, S.~Wang, T.~Guillod, M.~Luo, V.~Bansal, N.~K. Jha, Y.~Chen,
  C.~R. Sullivan, and M.~Chen, ``Why magnet: Quantifying the complexity of
  modeling power magnetic material characteristics,'' \emph{IEEE Transactions
  on Power Electronics}, 2023.

\bibitem{li2023magnet_2}
H.~Li, D.~Serrano, S.~Wang, and M.~Chen, ``Magnet-ai: Neural network as
  datasheet for magnetics modeling and material recommendation,'' \emph{IEEE
  Transactions on Power Electronics}, 2023.

\bibitem{grad}
E.~Polak, \emph{Optimization: algorithms and consistent approximations}.\hskip
  1em plus 0.5em minus 0.4em\relax Springer Science \& Business Media, 2012,
  vol. 124.

\bibitem{sa}
S.~Kirkpatrick, C.~D. Gelatt~Jr, and M.~P. Vecchi, ``Optimization by simulated
  annealing,'' \emph{science}, vol. 220, no. 4598, pp. 671--680, 1983.

\bibitem{pso1}
M.~R. Bonyadi and Z.~Michalewicz, ``Particle swarm optimization for single
  objective continuous space problems: A review,'' \emph{Evolutionary
  Computation}, vol.~25, no.~1, pp. 1--54, 2017.

\bibitem{pso2}
J.~Kennedy and R.~Eberhart, ``Particle swarm optimization,'' in
  \emph{Proceedings of ICNN'95 - International Conference on Neural Networks},
  vol.~4, 1995, pp. 1942--1948 vol.4.

\bibitem{evo}
K.~Price, R.~M. Storn, and J.~A. Lampinen, \emph{Differential evolution: a
  practical approach to global optimization}.\hskip 1em plus 0.5em minus
  0.4em\relax Springer Science \& Business Media, 2006.

\bibitem{ga}
A.~E. Eiben and J.~E. Smith, \emph{Introduction to evolutionary
  computing}.\hskip 1em plus 0.5em minus 0.4em\relax Springer-Verlag Berlin
  Heidelberg, 2015.

\bibitem{Chai2021}
\BIBentryALTinterwordspacing
J.~Chai, H.~Zeng, A.~Li, and E.~W. Ngai, ``Deep learning in computer vision: A
  critical review of emerging techniques and application scenarios,''
  \emph{Machine Learning with Applications}, vol.~6, p. 100134, 2021. [Online].
  Available:
  \url{https://www.sciencedirect.com/science/article/pii/S2666827021000670}
\BIBentrySTDinterwordspacing

\bibitem{Alam2020}
\BIBentryALTinterwordspacing
M.~Alam, M.~Samad, L.~Vidyaratne, A.~Glandon, and K.~Iftekharuddin, ``Survey on
  deep neural networks in speech and vision systems,'' \emph{Neurocomputing},
  vol. 417, pp. 302--321, 2020. [Online]. Available:
  \url{https://www.sciencedirect.com/science/article/pii/S0925231220311619}
\BIBentrySTDinterwordspacing

\bibitem{selvaratnam2021}
B.~Selvaratnam and R.~T. Koodali, ``Machine learning in experimental materials
  chemistry,'' \emph{Catalysis Today}, vol. 371, pp. 77--84, 2021.

\bibitem{voyles2017}
P.~M. Voyles, ``Informatics and data science in materials microscopy,''
  \emph{Current Opinion in Solid State and Materials Science}, vol.~21, no.~3,
  pp. 141--158, 2017.

\bibitem{schutt2019}
K.~T. Sch{\"u}tt, M.~Gastegger, A.~Tkatchenko, K.-R. M{\"u}ller, and R.~J.
  Maurer, ``Unifying machine learning and quantum chemistry with a deep neural
  network for molecular wavefunctions,'' \emph{Nature communications}, vol.~10,
  no.~1, p. 5024, 2019.

\bibitem{chen2019}
C.~Chen, W.~Ye, Y.~Zuo, C.~Zheng, and S.~P. Ong, ``Graph networks as a
  universal machine learning framework for molecules and crystals,''
  \emph{Chemistry of Materials}, vol.~31, no.~9, pp. 3564--3572, 2019.

\bibitem{Li_2022}
\BIBentryALTinterwordspacing
W.~Li, P.~Chen, B.~Xiong, G.~Liu, S.~Dou, Y.~Zhan, Z.~Zhu, T.~Chu, Y.~Li, and
  W.~Ma, ``Deep learning modeling strategy for material science: from natural
  materials to metamaterials,'' \emph{Journal of Physics: Materials}, vol.~5,
  no.~1, p. 014003, mar 2022. [Online]. Available:
  \url{https://dx.doi.org/10.1088/2515-7639/ac5914}
\BIBentrySTDinterwordspacing

\bibitem{Fuhr2022}
\BIBentryALTinterwordspacing
A.~S. Fuhr and B.~G. Sumpter, ``Deep generative models for materials discovery
  and machine learning-accelerated innovation,'' \emph{Frontiers in Materials},
  vol.~9, 2022. [Online]. Available:
  \url{https://www.frontiersin.org/articles/10.3389/fmats.2022.865270}
\BIBentrySTDinterwordspacing

\bibitem{Choudhary2022}
\BIBentryALTinterwordspacing
K.~Choudhary, B.~DeCost, C.~Chen, A.~Jain, F.~Tavazza, R.~Cohn, C.~W. Park,
  A.~Choudhary, A.~Agrawal, S.~J.~L. Billinge, E.~Holm, S.~P. Ong, and
  C.~Wolverton, ``Recent advances and applications of deep learning methods in
  materials science,'' \emph{npj Computational Materials}, vol.~8, no.~1,
  p.~59, 2022. [Online]. Available:
  \url{https://doi.org/10.1038/s41524-022-00734-6}
\BIBentrySTDinterwordspacing

\bibitem{Bonatti2021}
\BIBentryALTinterwordspacing
C.~Bonatti and D.~Mohr, ``One for all: Universal material model based on
  minimal state-space neural networks,'' \emph{Science Advances}, vol.~7,
  no.~26, p. eabf3658, 2021. [Online]. Available:
  \url{https://www.science.org/doi/abs/10.1126/sciadv.abf3658}
\BIBentrySTDinterwordspacing

\bibitem{damewood2023}
J.~Damewood, J.~Karaguesian, J.~R. Lunger, A.~R. Tan, M.~Xie, J.~Peng, and
  R.~G{\'o}mez-Bombarelli, ``Representations of materials for machine
  learning,'' \emph{Annual Review of Materials Research}, vol.~53, 2023.

\bibitem{corednn1}
H.~Saliah, D.~Lowther, and B.~Forghani, ``A neural network model of magnetic
  hysteresis for computational magnetics,'' \emph{IEEE Transactions on
  Magnetics}, vol.~33, no.~5, pp. 4146--4148, 1997.

\bibitem{corednn2}
C.~Serpico and C.~Visone, ``Magnetic hysteresis modeling via feed-forward
  neural networks,'' \emph{IEEE Transactions on Magnetics}, vol.~34, no.~3, pp.
  623--628, 1998.

\bibitem{corednn3}
H.~Saliah, D.~Lowther, and B.~Forghani, ``Modeling magnetic materials using
  artificial neural networks,'' \emph{IEEE Transactions on Magnetics}, vol.~34,
  no.~5, pp. 3056--3059, 1998.

\bibitem{corednn4}
\BIBentryALTinterwordspacing
F.~Sixdenier, R.~Scorretti, R.~Marion, and L.~Morel, ``Quasistatic hysteresis
  modeling with feed-forward neural networks: Influence of the last but one
  extreme values,'' \emph{Journal of Magnetism and Magnetic Materials}, vol.
  320, no.~20, pp. e992--e996, 2008, proceedings of the 18th International
  Symposium on Soft Magnetic Materials. [Online]. Available:
  \url{https://www.sciencedirect.com/science/article/pii/S030488530800543X}
\BIBentrySTDinterwordspacing

\bibitem{corednn5}
Z.~Zhao, F.~Liu, S.~L. Ho, W.~N. Fu, and W.~Yan, ``Modeling magnetic hysteresis
  under dc-biased magnetization using the neural network,'' \emph{IEEE
  Transactions on Magnetics}, vol.~45, no.~10, pp. 3958--3961, 2009.

\bibitem{corednn6}
F.~Riganti~Fulginei and A.~Salvini, ``Neural network approach for modelling
  hysteretic magnetic materials under distorted excitations,'' \emph{IEEE
  Transactions on Magnetics}, vol.~48, no.~2, pp. 307--310, 2012.

\bibitem{corednn7}
\BIBentryALTinterwordspacing
S.~{Quondam Antonio}, F.~{Riganti Fulginei}, A.~Laudani, A.~Faba, and
  E.~Cardelli, ``An effective neural network approach to reproduce magnetic
  hysteresis in electrical steel under arbitrary excitation waveforms,''
  \emph{Journal of Magnetism and Magnetic Materials}, vol. 528, p. 167735,
  2021. [Online]. Available:
  \url{https://www.sciencedirect.com/science/article/pii/S0304885321000111}
\BIBentrySTDinterwordspacing

\bibitem{corernn}
\BIBentryALTinterwordspacing
C.~Grech, M.~Buzio, M.~Pentella, and N.~Sammut, ``Dynamic ferromagnetic
  hysteresis modelling using a preisach-recurrent neural network model,''
  \emph{Materials}, vol.~13, no.~11, 2020. [Online]. Available:
  \url{https://www.mdpi.com/1996-1944/13/11/2561}
\BIBentrySTDinterwordspacing

\bibitem{magnet1}
H.~Li, D.~Serrano, S.~Wang, and M.~Chen, ``Magnet-ai: Neural network as
  datasheet for magnetics modeling and material recommendation,'' \emph{IEEE
  Transactions on Power Electronics}, vol.~38, no.~12, pp. 15\,854--15\,869,
  2023.

\bibitem{magnet2}
H.~Li, D.~Serrano, T.~Guillod, S.~Wang, E.~Dogariu, A.~Nadler, M.~Luo,
  V.~Bansal, N.~K. Jha, Y.~Chen, C.~R. Sullivan, and M.~Chen, ``How magnet:
  Machine learning framework for modeling power magnetic material
  characteristics,'' \emph{IEEE Transactions on Power Electronics}, vol.~38,
  no.~12, pp. 15\,829--15\,853, 2023.

\bibitem{ferch2013application}
M.~Ferch, ``Application overview of nanocrystalline inductive components in
  today’s power electronic systems,'' in \emph{Proc. Soft Magn. Mater. Conf},
  no. A1-01, 2013.

\bibitem{riechert2012mitigation}
U.~Riechert, M.~B{\"o}sch, M.~Szewczyk, W.~Piasecki, J.~Smajic, A.~Shoory,
  S.~Burow, and S.~Tenbohlen, ``Mitigation of very fast transient overvoltages
  in gas insulated uhv substations,'' \emph{Proc. CIGRE}, 2012.

\bibitem{JA1984}
\BIBentryALTinterwordspacing
D.~C. Jiles and D.~L. Atherton, ``{Theory of ferromagnetic hysteresis
  (invited)},'' \emph{Journal of Applied Physics}, vol.~55, no.~6, pp.
  2115--2120, 03 1984. [Online]. Available:
  \url{https://doi.org/10.1063/1.333582}
\BIBentrySTDinterwordspacing

\bibitem{Liorzou2000}
F.~Liorzou, B.~Phelps, and D.~Atherton, ``Macroscopic models of
  magnetization,'' \emph{IEEE Transactions on Magnetics}, vol.~36, no.~2, pp.
  418--428, 2000.

\bibitem{Perkkio2018}
L.~Perkkiö, B.~Upadhaya, A.~Hannukainen, and P.~Rasilo, ``Stable adaptive
  method to solve fem coupled with jiles–atherton hysteresis model,''
  \emph{IEEE Transactions on Magnetics}, vol.~54, no.~2, pp. 1--8, 2018.

\bibitem{Hoffmann2016}
K.~Hoffmann, J.~P. Assumpção~Bastos, J.~V. Leite, N.~Sadowski, and
  F.~Barbosa, ``An accurate vector jiles-atherton model for improving the fem
  convergence,'' in \emph{2016 IEEE Conference on Electromagnetic Field
  Computation (CEFC)}, 2016, pp. 1--1.

\bibitem{leecore}
M.~Chen, M.~Araghchini, K.~K. Afridi, J.~H. Lang, C.~R. Sullivan, and D.~J.
  Perreault, ``A systematic approach to modeling impedances and current
  distribution in planar magnetics,'' \emph{IEEE Transactions on Power
  Electronics}, vol.~31, no.~1, pp. 560--580, 2016.

\bibitem{spice}
H.~Dommel and W.~Meyer, ``Computation of electromagnetic transients,''
  \emph{Proceedings of the IEEE}, vol.~62, no.~7, pp. 983--993, 1974.

\bibitem{dommel1974computation}
H.~W. Dommel and W.~S. Meyer, ``Computation of electromagnetic transients,''
  \emph{Proceedings of the IEEE}, vol.~62, no.~7, pp. 983--993, 1974.

\bibitem{abb2016}
M.~Szewczyk, K.~Kutorasiński, J.~Pawłowski, W.~Piasecki, and M.~Florkowski,
  ``Advanced modeling of magnetic cores for damping of high-frequency power
  system transients,'' \emph{IEEE Transactions on Power Delivery}, vol.~31,
  no.~5, pp. 2431--2439, 2016.

\bibitem{bib:NMMcircuitSimulation}
K.~Kutorasi{\'n}ski, J.~Paw{\l}owski, P.~Leszczy{\'n}ski, and M.~Szewczyk,
  ``Nonlinear modeling of magnetic materials for circuit simulations,''
  \emph{Scientific Reports}, vol.~13, no.~1, p. 17178, 2023.

\bibitem{PawlowskiJ2022Mnmo}
J.~Pawłowski, K.~Kutorasiński, and M.~Szewczyk,
  ``\BIBforeignlanguage{eng}{Multifrequency nonlinear model of magnetic
  material with artificial intelligence optimization},''
  \emph{\BIBforeignlanguage{eng}{Scientific reports}}, vol.~12, no.~1, pp.
  19\,784--19\,784, 2022.

\bibitem{bib:MeasurementMR}
\BIBentryALTinterwordspacing
K.~Kutorasiński, M.~Szewczyk, M.~Molas, and J.~Pawłowski, ``Measuring
  impedance frequency characteristics of magnetic rings with dc-bias current,''
  \emph{ISA Transactions}, 2023. [Online]. Available:
  \url{https://www.sciencedirect.com/science/article/pii/S0019057823003968}
\BIBentrySTDinterwordspacing

\bibitem{nikolenko2021}
S.~I. Nikolenko, \emph{Synthetic data for deep learning (Springer Optimization
  and Its Applications)}.\hskip 1em plus 0.5em minus 0.4em\relax Springer,
  2021, vol. 174.

\bibitem{zeiler2014visualizing}
M.~D. Zeiler and R.~Fergus, ``Visualizing and understanding convolutional
  networks,'' in \emph{European conference on computer vision}.\hskip 1em plus
  0.5em minus 0.4em\relax Springer, 2014, pp. 818--833.

\bibitem{Yu2016dilated}
F.~Yu and V.~Koltun, ``Multi-scale context aggregation by dilated
  convolutions,'' in \emph{International Conference on Learning Representations
  (ICLR)}, May 2016.

\bibitem{attentionpaper}
\BIBentryALTinterwordspacing
A.~Vaswani, N.~Shazeer, N.~Parmar, J.~Uszkoreit, L.~Jones, A.~N. Gomez, L.~u.
  Kaiser, and I.~Polosukhin, ``Attention is all you need,'' in \emph{Advances
  in Neural Information Processing Systems}, I.~Guyon, U.~V. Luxburg,
  S.~Bengio, H.~Wallach, R.~Fergus, S.~Vishwanathan, and R.~Garnett, Eds.,
  vol.~30.\hskip 1em plus 0.5em minus 0.4em\relax Curran Associates, Inc.,
  2017. [Online]. Available:
  \url{https://proceedings.neurips.cc/paper_files/paper/2017/file/3f5ee243547dee91fbd053c1c4a845aa-Paper.pdf}
\BIBentrySTDinterwordspacing

\bibitem{FaceNet}
\BIBentryALTinterwordspacing
F.~Schroff, D.~Kalenichenko, and J.~Philbin, ``Facenet: A unified embedding for
  face recognition and clustering,'' \emph{2015 IEEE Conference on Computer
  Vision and Pattern Recognition (CVPR)}, Jun 2015. [Online]. Available:
  \url{http://dx.doi.org/10.1109/CVPR.2015.7298682}
\BIBentrySTDinterwordspacing

\bibitem{gan}
\BIBentryALTinterwordspacing
I.~Goodfellow, J.~Pouget-Abadie, M.~Mirza, B.~Xu, D.~Warde-Farley, S.~Ozair,
  A.~Courville, and Y.~Bengio, ``Generative adversarial networks,''
  \emph{Commun. ACM}, vol.~63, no.~11, p. 139–144, oct 2020. [Online].
  Available: \url{https://doi.org/10.1145/3422622}
\BIBentrySTDinterwordspacing

\bibitem{styletransfer}
J.~Johnson, A.~Alahi, and L.~Fei-Fei, ``Perceptual losses for real-time style
  transfer and super-resolution,'' in \emph{Computer Vision -- ECCV 2016},
  B.~Leibe, J.~Matas, N.~Sebe, and M.~Welling, Eds.\hskip 1em plus 0.5em minus
  0.4em\relax Cham: Springer International Publishing, 2016, pp. 694--711.

\bibitem{szewczyk2014identification}
M.~Szewczyk, J.~Pawlowski, K.~Kutorasinski, S.~Burow, S.~Tenbohlen, and
  W.~Piasecki, ``Identification of rational function in s-domain describing a
  magnetic material frequency characteristics,'' in \emph{9th IET International
  Conference on Computation in Electromagnetics (CEM 2014)}.\hskip 1em plus
  0.5em minus 0.4em\relax IET, 2014, pp. 1--2.

\bibitem{gustavsen1999rational}
B.~Gustavsen and A.~Semlyen, ``Rational approximation of frequency domain
  responses by vector fitting,'' \emph{IEEE Transactions on power delivery},
  vol.~14, no.~3, pp. 1052--1061, 1999.

\bibitem{de2008single}
L.~De~Tommasi, D.~Deschrijver, and T.~Dhaene, ``Single-input-single-output
  passive macromodeling via positive fractions vector fitting,'' in \emph{2008
  12th IEEE Workshop on Signal Propagation on Interconnects}.\hskip 1em plus
  0.5em minus 0.4em\relax IEEE, 2008, pp. 1--2.

\bibitem{paszke2019pytorch}
A.~Paszke, S.~Gross, F.~Massa, A.~Lerer, J.~Bradbury, G.~Chanan, T.~Killeen,
  Z.~Lin, N.~Gimelshein, L.~Antiga \emph{et~al.}, ``Pytorch: An imperative
  style, high-performance deep learning library,'' \emph{Advances in neural
  information processing systems}, vol.~32, 2019.

\bibitem{magnetec_datasheet_1}
``Product specification for inductive components: M-676 core datasheet,''
  \emph{MAGNETEC GmbH}, Jan, 2020.

\end{thebibliography}

\end{document}